\documentclass[preprint,superscriptaddress,amsmath,amssymb]{revtex4}
\usepackage{graphicx}
\usepackage{color}
\usepackage{enumerate}
\usepackage[caption=false]{subfig}   

\usepackage[linktocpage=true, pdfborder={0 0 0},pdfpagelabels,plainpages=false]{hyperref}

\newcommand {\norm}[1] {\left| #1 \right|}
\newcommand {\klamm}[1] {\left( #1 \right)}    
\newcommand {\sklamm}[1] {\left[ #1 \right]}   
\newcommand {\eklamm}[1] {\left \langle #1 \right \rangle}   

\newcommand {\vek}[1]{\mathbf{\boldsymbol{#1}}}  

\newcommand{\exppsi}[0]{\eklamm{\norm{\Psi(\vek{r})}^2}_E}

\newcommand{\bw}[1]{bw_}

\begin{document}

\title{Intensity distribution of non-linear scattering states}

\author{Timo Hartmann}
\author{Juan-Diego Urbina}
\author{Klaus Richter}
\affiliation{Institut f\"ur Theoretische Physik, Universit\"at Regensburg,
  93040 Regensburg, Germany}
\author{Peter Schlagheck}
\affiliation{D\'epartement de Physique,
Universit\'e de Li\`ege, 4000 Li\`ege, Belgium}

\begin{abstract}

We investigate the interplay between coherent effects characteristic of the propagation
of linear waves, the non-linear effects due to interactions, and the
quantum manifestations of classical chaos due to geometrical
confinement, as they arise in the context of the transport of 
Bose-Einstein condensates.
We specifically show that, extending standard methods for non-interacting 
systems, {the body of the statistical distribution of intensities for} scattering states solving the 
Gross-Pitaevskii equation {is} very well described by a local Gaussian 
ansatz with a position-dependent variance.
We propose a semiclassical approach based on interfering classical paths 
to fix the single parameter describing the universal deviations from a 
global Gaussian distribution. {Being tail effects, rare events like rogue waves characteristic of non-linear field equations do not affect our results}. 

\end{abstract}

\maketitle

\section{Introduction}

The progress on the experimental preparation and manipulation of 
interacting Bose-Einstein condensates has given a strong boost to the 
study of non-linear wave equations that account for the
effect of interactions within the condensate in the framework of 
a mean-field approximation.
Particularly promising cold-atom experiments in the context of transport 
physics include the realization of guided atom lasers 
\cite{GueO06PRL,CouO08EPL,FabO11PRL,GatO11PRL}, of arbitrarily 
shaped confinement potentials for cold atoms 
\cite{MilO01PRL,FriO01PRL,1367-2630-11-4-043030}, as well as of
artificial gauge fields that break the time-reversal invariance for 
neutral atoms \cite{RevModPhys.83.1523,PhysRevA.71.053614}.
This makes it now feasible to experimentally explore the coherent 
transport of Bose-Einstein condensates through various mesoscopic structures 
that can possibly be modeled by billiard configurations.

An interesting question that rises in this context 
is how the presence of the atom-atom interaction within the coherent 
matter waves affects interference effects well that are established for
non-interacting systems.
Indeed, previous theoretical studies have focused on the question how 
coherent backscattering in disordered potentials is modified by the 
presence of the atom-atom interaction~\cite{PhysRevLett.101.020603}.
These studies were recently complemented by our investigations on
weak localization in the nonlinear transport through ballistic scattering 
geometries that exhibit chaotic dynamics \cite{weakloc}.
While a semiclassical analysis of this nonlinear scattering problem predicted
a dephasing of the interference phenomenon that gives rise to coherent 
backscattering, signatures for weak antilocalization were obtained in the
numerically computed reflection and transmission probabilities \cite{weakloc}.
This effect was attributed to the specific role of non-universal short-path
contributions, in particular to self-retraced paths the presence of which
gives rise to a reduction of coherent backscattering as compared to the
universal prediction.

In the present work, we consider the same scenario as in Ref.~\cite{weakloc},
i.e., the quasi-stationary transport  of bosonic matter waves 
through two-dimensional ballistic scattering geometries that exhibit 
chaotic classical dynamics.
In contrast to Ref.~\cite{weakloc}, however, we focus here not on
transport observables such as the reflection and transmission probabilities
through the billiard, but rather on the intensity distributions of 
stationary scattering states within the billiard.
These intensity distributions are to be compared with the theoretical 
predictions that are obtained from the Random Wave Model 
(RWM) \cite{0305-4470-10-12-016,PhysRevLett.97.214101,RW1,RW2}, which, 
in the linear case, represents probably the most powerful approach to 
describe the universal spatial correlations of eigenstates arising from 
the classical chaotic behavior due to the presence of a spatial
confinement. 
A most natural question that arises here is then to which extent 
the basic assumptions behind this model can also be used to 
describe possible universal spatial fluctuations in collective coherent
matter waves that exhibit a weak atom-atom interaction.
Within a mean-field semiclassical description, such matter waves are well
described by the Gross-Pitaevskii equation~\cite{DalO99RMP} in which
the presence of interaction is accounted for by means of a 
non-linear interaction potential.
This equation is the starting point of our calculations, 
both on the numerical and on the analytical side. 

{It is important to mention that rare effects due to the nonlinearity of the wave equation like rogue waves \cite{RoW} or due to the presence of disorder, like branching \cite{B1}, will certainly affect the tails of the intensity distribution, and such effects are in principle outside the reach of our approach. Therefore, we will rstrict ourselves to the body of the distribution, where rare events need not to be considered.}   

The paper is organized as follows.
We describe in Section \ref{sec:ssgpe} the scattering configuration under
consideration as well as the main observable to be discussed in this work.
In Section \ref{sec:inten}, we present a semiclassical theory of the
intensity distribution in this nonlinear system, which is based on the
Gaussian hypothesis as well as on the semiclassical theory of coherent
backscattering in nonlinear systems.
The predictions obtained by this semiclassical theory will be compared with
the numerical results at the end of Section \ref{sec:inten}, followed by
a discussion in Section \ref{sec:conc}.

\section{Stationary scattering states of the Gross-Pitaevskii equation}

\label{sec:ssgpe}

For our simulations, we use the inhomogeneous two-dimensional 
Gross-Pitaevskii equation 
\begin{equation}
\label{eq:gp_tp}
i \hbar \frac{\partial}{\partial t}\Psi(\vek{r},t) = H \Psi(\vek{r},t) + g(\vek{r}) \frac{\hbar^2}{m} |\Psi(\vek{r},t)|^2 \Psi(\vek{r},t) + S(\vek{r}) e^{-i t \mu / \hbar}
\end{equation}
where we have introduced the single particle Hamiltonian
\begin{equation}
H=\frac{1}{2m}\sklamm{-i\hbar \vek{\nabla} - q \vek{A}(\vek{r})}^2 + V(\vek{r})
\end{equation}
{with the billiard potential $V(\vek{r})$.
This Gross-Pitaevskii equation contains} a source term
\begin{equation}
S(\mathbf{r}) = S_0 \chi_i(y) \delta(x - x_L)\;
\end{equation}
which models the injection of atoms {in a Bose-Einstein condensate acting as a reservoir} with the chemical potential $\mu$ 
into the scattering system {\cite{ErnPauSch10PRA}}.
$\chi_i(y)$ is a transverse eigenmode of the {incident} lead and 
$S_0$ controls the current {that} is injected into the billiard.

The non-linear potential term 
$g(\vek{r}) \frac{\hbar^2}{m} |\Psi(\vek{r},t)|^2 \Psi(\vek{r},t)$ 
describes atom-atom scattering events.
Assuming that the degree of motion for the third spatial dimension is 
 frozen out{, e.g.}\ by applying a
harmonic confinement potential 
in this direction, we obtain the effective two-dimensional 
interaction strength as
$g(\mathbf{r}) = {2} \sqrt{2 \pi} a_s/
\sqrt{\hbar/[m\omega_\perp(\mathbf{r})]}$ 
where $a_s$ is the s-wave scattering length of the atomic species
under consideration and $\omega_\perp$ is the confinement frequency in the
third spatial dimension. 
A spatial variation $\omega_\perp \equiv \omega_\perp(\mathbf{r})$ of this 
confinement will then naturally induce a corresponding variation in 
$g\equiv g(\mathbf{r})$.
We shall, in the following, consider an effective interaction strength
$g(\mathbf{r})$ that is homogeneous within the billiard and vanishes in the
attached leads.
In a similar manner, we shall also assume that the artificial gauge field
is given by $\mathbf{A}(\mathbf{r}) = \frac{1}{2} B 
\mathbf{e}_\perp \times \mathbf{r}$
(with $\mathbf{e}_\perp$ the unit vector in the third spatial dimension),
with an effective ``magnetic field'' strength $B$ that is constant within 
the billiard and vanishes in the leads.

\begin{figure}[tb]
\includegraphics[width=0.49\linewidth]{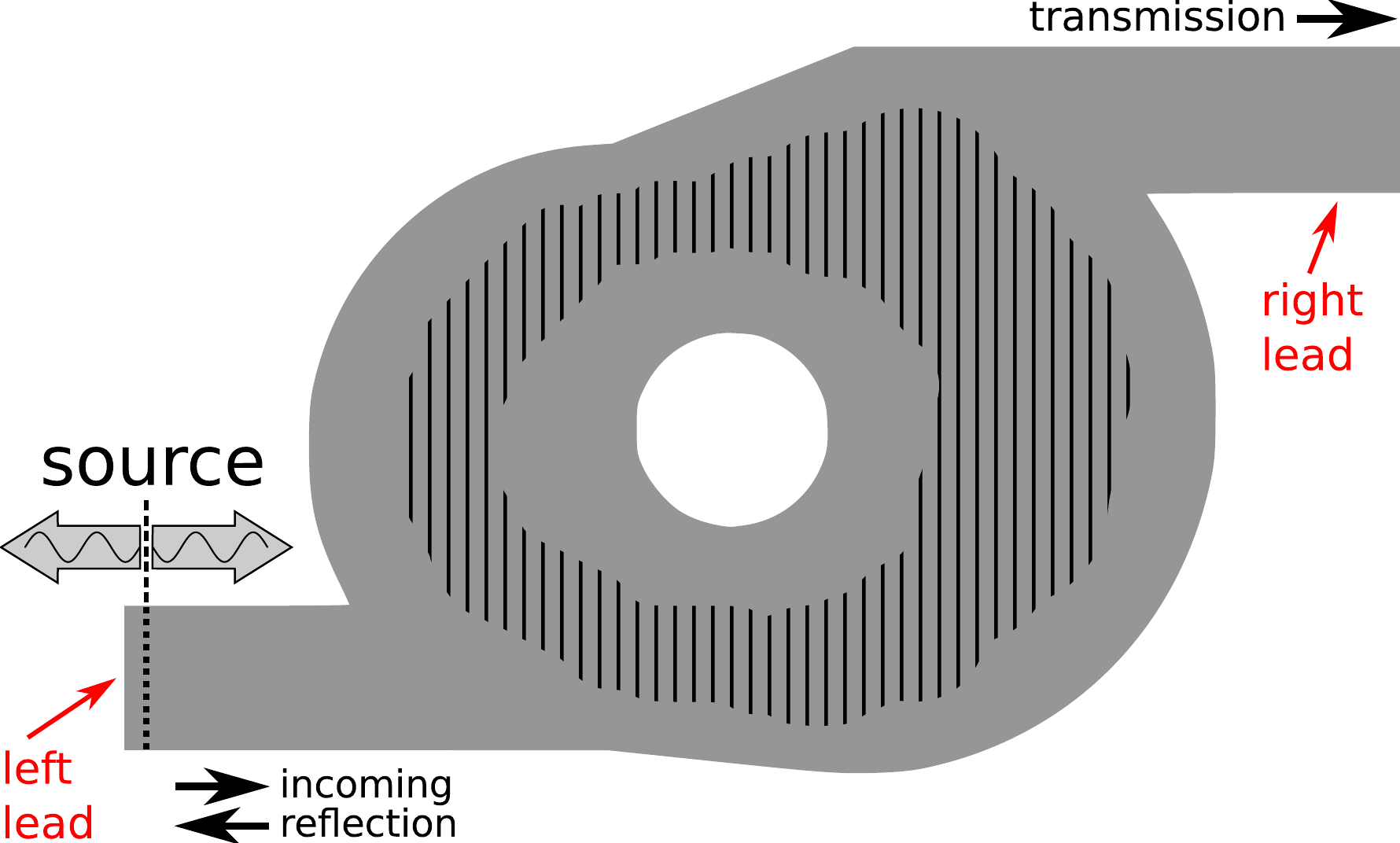}
\hfill
\includegraphics[width=0.49\linewidth]{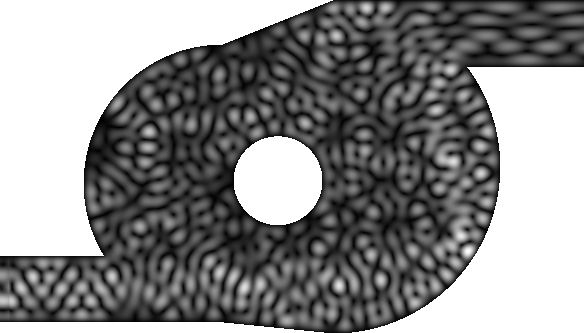}
\caption{\label{fig:billiard}
The shape of the billiard used in this work together with the density of a 
typical stationary scattering state. 
The hatched area in the left figure marks the
region used for calculating the intensity distribution {(adapted from Ref.~\cite{weakloc})}.
}
\end{figure}

The billiard geometry considered in this work is shown in 
Fig.~\ref{fig:billiard}. 
It is characterized by {the billiard} area $\Omega$ and 
{the} typical energy $E_0$ of the incident matter-wave beam.
Using these quantities{,} we can define a time scale $\hbar/E_0$,
a length scale $k_0^{-1}$ {with $E_0\equiv\hbar^2k^2_0/(2 m)$},
and a scale {$B_0\equiv 2\pi \hbar/(q\Omega)$} (the flux quantum) 
for the magnetic field.
All quantities in this work will be measured in these units.
The area of the system is determined {as} $k_0\Omega^{1/2}=81.2$. 
Two leads are attached to the billiard{,} which transforms it into an 
open scattering system.
The width of the leads is given by $W=5.4 \, \pi/k_0$, 
which means that five channels are open in each of the leads.

In order to calculate the stationary scattering states within this
configuration, we insert the ansatz
\begin{equation}
\Psi(\vek{r},t)=\Psi(\vek{r}) e^{-i t \mu / \hbar}
\end{equation}
into Eq.~\eqref{eq:gp_tp}. This yields the self-consistent non-linear equation
\begin{equation}
\label{eq:nl_scatter}
\sklamm{\mu-H-g(\vek{r})\frac{\hbar^2}{m}|\Psi(\vek{r})|^2}\Psi(\vek{r})=S(\vek{r})
\end{equation}
for the stationary scattering state.
The amplitude of the source term is fixed such that in incident current of 
$j_{in}=1\, E_0/\hbar$ is generated. 
Varying $j_{in}$ provides yet another way to effectively change the 
interaction strength $g$, as Eq.~\eqref{eq:nl_scatter} is invariant 
under the scaling 
$(g,j_{in},\Psi) \mapsto (g \, \eta^{-2}, j_{in} \, \eta^2, \Psi \, \eta)$
(for $\eta\in\mathbb{R}$).

The non-linear scattering problem Eq.~\eqref{eq:nl_scatter} is now solved 
using the methods {described in Appendix} \ref{sec:numerics}{.
We performed computations for} 50 different values of the energy $\mu$ 
(all in the energy range $0.93 \, E_0 \dots 1.18 \, E_0 $ 
where five lead channels are open),
for 25 different positions of the spherical obstacle in the centre 
of the billiard, and for the five different lead channels.
The {thereby} obtained stationary scattering states 
$\Psi(\vek{r})$ are now used to determine the intensity distribution, 
i.e.\ the probability distribution of $|\Psi(\vek{r})|^2$, 
and its mean value.
Only the points inside the marked region in Fig.~\ref{fig:billiard} were used.
Points in the vicinity of a boundary have to be avoided as explained in 
Sec.~\ref{sec:inten}.

\begin{figure}[tb]
\includegraphics[width=0.48\linewidth]{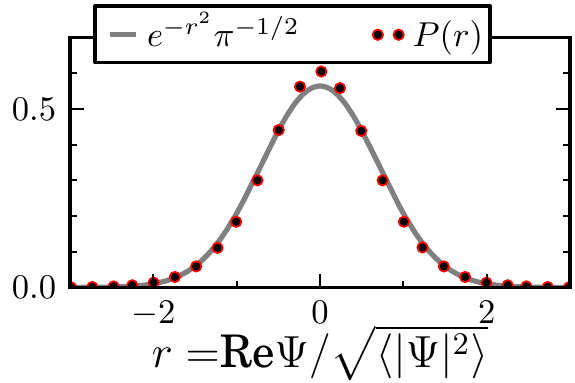}
 \hfill
\includegraphics[width=0.48\linewidth]{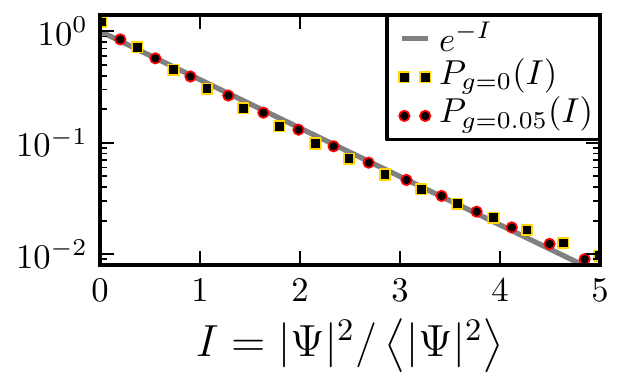}
\caption{\label{fig:gauss}
Left panel: numerical probability distribution (dots) of the real part 
of the wavefunction for $B=0$ and $g=0$, which agrees very well with a
Gaussian distribution (solid line). 
The same holds for the imaginary part.
The right panel compares the numerically obtained intensity 
distributions $P_g(I)$ for $g=0$ and $g=0.05$ ($B=0$ in both cases) 
with the {Porter-Thomas} distribution $e^{-I}$.
Note the tiny but systematic deviations from the Porter-Thomas law, which are
highlighted in Fig.~\ref{fig:dist}.
}
\end{figure}

The left panel of Fig.~\ref{fig:gauss} shows the probability distribution
for obtaining a given real part of the scattering wavefunction (which is the 
same as for the imaginary part) within the marked region of the billiard
in the linear ($g=0$) and time-reversal invariant ($B=0$) case.
We find a very good agreement with a Gaussian distribution.
Consequently, the intensity 
$I \equiv |\Psi|^2 / \langle |\Psi|^2 \rangle$ is distributed according
to a {Porter-Thomas} law $P(I) \simeq e^{-I}$, as is confirmed in the right panel
of Fig.~\ref{fig:gauss}.
There are tiny but systematic deviations from the {Porter-Thomas} law which slightly
underestimates the actual intensity distribution near $I=0$ (as is also seen
in the left panel of Fig.~\ref{fig:gauss}) as well as for very large 
intensities $I \gtrsim 5$, and overestimates it in between for 
$1 \lesssim I \lesssim 3$.

\begin{figure}[tb]
\includegraphics[width=0.49\linewidth]{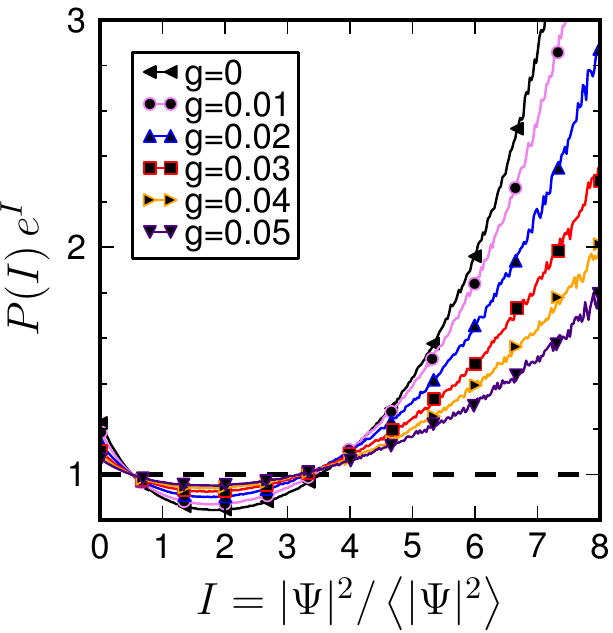}
\hfill
\includegraphics[width=0.49\linewidth]{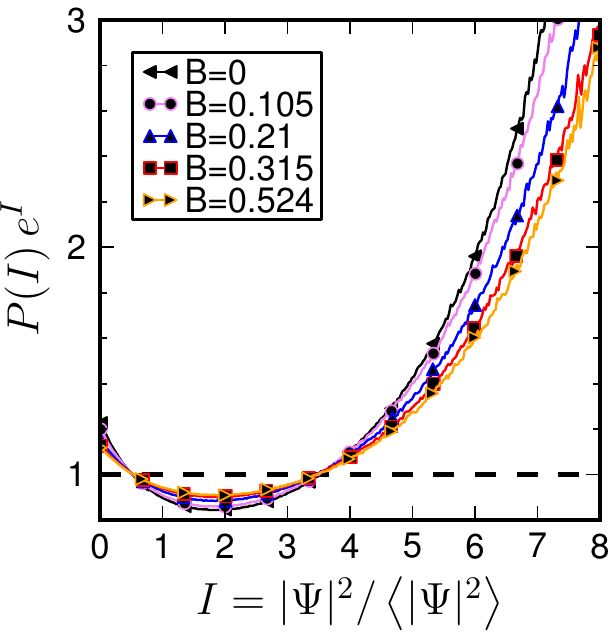}
\caption{\label{fig:dist}
Deviation of the intensity distributions from the {Poster-Thomas} law 
for several values of the interaction strength $g$ 
(left panel, with $B=0$) and of the magnetic field $B$ 
in units of $B_0$ (right panel, for $g=0$)}.
\end{figure}

To highlight these deviations, we plot in Fig.~\ref{fig:dist}
$P(I) e^I$ as a function of the intensity $I$, for various values of the
nonlinearity $g$ and the magnetic field strength $B$.
A parabolic behaviour with a minimum at $I = 2$ is found.
The prefactor of this parabolic scaling is reduced with increasing $g$.
This appears natural as a weak repulsive interaction between the atoms is 
generally expected to give rise to a flattening of the density distribution,
leading, in particular, to a significant reduction of intensity maxima,
in order to minimize the interaction energy within the condensate
(see also Ref.~\cite{SchO07N} for an analogous phenomenology in nonlinear 
optics in the presence of a defocusing nonlinearity).
Indeed, similar findings were obtained for the quasi-stationary transport of
Bose-Einstein condensates through two-dimensional disorder potentials
\cite{Hartung}, for which is was found that the parabolic scaling of 
$P(I) e^I$ with the intensity $I$ could even become inverted at stronger 
nonlinearities $g$.
The dependence of the parabolic scaling with the magnetic field $B$, on the
other hand, is related to \emph{coherent backscattering}, for which we shall 
develop a semiclassical theory in the following section.

\section{The semiclassical approach to the intensity distribution}
\label{sec:inten}

In a first step, and following the now standard approach to describe
the statistical properties of eigenfunctions in non-interacting and
classically chaotic billiard systems \cite{0305-4470-10-12-016}, 
we shall make the fundamental assumption that 
\textit{scattering eigenstates of the non-linear Schr\"odinger equation 
share the same correlations {as} an ensemble of Gaussian Random Fields} 
(see the left panel of Fig.~\ref{fig:gauss}).
This assumption leads to a Poster-Thomas distribution $P(I)=e^{-I}$
for the normalized intensity $I=|\Psi|^2/\eklamm{|\Psi|^2}$ 
(see Eq.~\eqref{eq:poisson} below) which, as discussed above, 
is supported by our numerical findings, as seen in the right panel 
of Fig.~\ref{fig:gauss}.
The presence of a weak interaction does not change the excellent agreement
of the numerical data with a Porter-Thomas profile.

Knowing that the general features of the distribution of
intensities for nonlinear waves are well described by a Porter-Thomas
distribution, we now ask whether the deviations observed in
Fig.~\ref{fig:dist} have also such universal character. 
Once again, the guiding principle will be linear case, 
where deviations from the body of the distribution are consistent with
a Gaussian random field with a variance that smoothly depends on
the local position.
This consideration leads to an universal form of the
deviations given by a Laguerre polynomial, which therefore depend 
only on a single parameter \cite{PhysRevLett.97.214101}. 
 Fig.~\ref{fig:dist} shows  how this property of the 
non-interacting case takes over perfectly when interactions are present.

The final step will be the explicit calculation of the coefficient in
front of the polynomial corrections, and in particular its dependence
on the strength of the interaction and of the magnetic field. 
Here we shall  assume that a basic property of  scattering 
states in the linear case, namely that their average intensity over energy and
channels is related with the imaginary part of the full Green function, 
holds approximately  in the presence of interactions as well. 
Assuming ergodicity within the billiard and utilizing the semiclassical 
approach presented in Ref.~\cite{weakloc}, we obtain an explicit 
expression for the variation of the polynomial prefactor with the magnetic 
field strength for various values of the nonlinearity.

\subsection{The local Gaussian approach}
  
The calculation of the intensity distribution uses the values of 
$|\Psi(\vek{r})|^2$ at many different positions, incoming channels, 
and energies. Thus, both an energy and a position average is involved. 
Motivated by the idea that for fixed position $\vek{r}$, the average 
intensity over energy and channels $\exppsi$ is itself a smooth function of
$\vek{r}$, the double averaging procedure is now split apart.

We start therefore by assuming a position-dependent Gaussian
distribution 
\begin{equation}
{P_{\vek{r}}\klamm{\Psi_{r},\Psi_{i}}=\frac{1}{\pi\exppsi} 
\exp\sklamm{-\frac{\Psi_{r}^2 + \Psi_{i}^2}{\exppsi}}}
\end{equation}
for the real and the imaginary part of the wave 
function ($\Psi{\equiv}\Psi_{r}+i \Psi_{i}$) at a fixed point $\vek{r}$,
where $\exppsi$ denotes the energy and channel average 
of the intensity. 
For non-interacting systems with chaotic classical dynamics, such a
local Gaussian distribution is a rigorous consequence of the Random Wave
Model \cite{stoeckmann@chaos}, and the possible universality of the 
deviations from the fully homogeneous case, i.e.\ from the case
that $\exppsi$ is independent of the position $\vek{r}$, are encoded in 
$\exppsi$ (see Ref.~\cite{PhysRevLett.97.214101}). 
At a boundary the
wavefunction vanishes and thus $\exppsi$ vanishes there, too. 
Such boundary effects can also be incorporated in our approach.
In this work, however, we shall restrict our study to the bulk, 
and therefore points in the vicinity of a boundary will be avoided.
The distribution for the intensity ${\rho \equiv} \norm{\Psi}^2$ is now
calculated as 
\begin{eqnarray}
P_{\vek{r}}({\rho}) & = & \int_{-\infty}^{+\infty} {\int_{-\infty}^{+\infty}}
P_{\vek{r}}(\Psi_{r},\Psi_{i}) \; \delta({\rho}-{\Psi_r^2 - \Psi_i^2}) 
\; d\Psi_r d\Psi_i
\nonumber \\ && 
\nonumber \\ & = &
\frac{1}{\exppsi}  \exp\sklamm{-\frac{{\rho}}{\exppsi}} \,{,}
\label{eq:poisson}
\end{eqnarray}
which is a {\it local} Porter-Thomas distribution for ${\rho}$.

\begin{figure}[tb]
 \begin{minipage}[c]{0.48\linewidth}
  \includegraphics[width=1\linewidth]{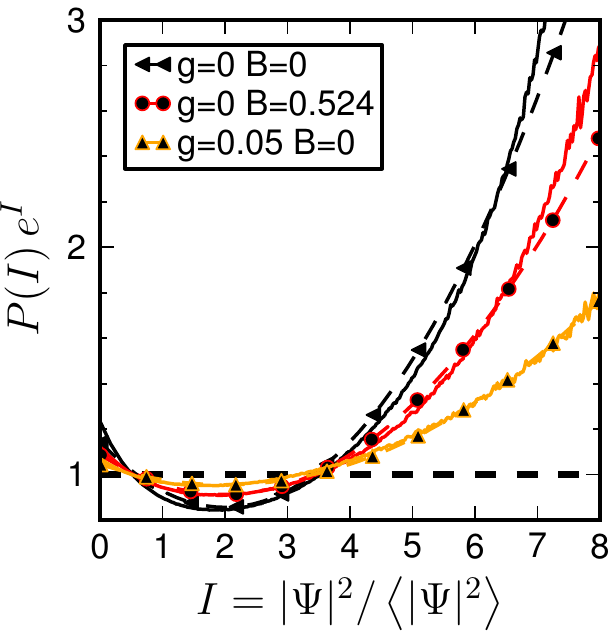}%
 \end{minipage}\hfill
 \begin{minipage}[c]{0.48\linewidth}
\caption{\label{fig:dist_fit}
{Comparison of the} numerically obtained intensity distributions with 
Eq.~\eqref{eq:idist}.
The unknown parameter $\beta$ is determined by fitting Eq.~\eqref{eq:idist} 
(shown as dashed lines and marker symbols) to the numerical data 
(solid lines). 
The fitting is done in the range $1 \leq I \leq 7$.
}
\end{minipage}
\end{figure}

We now proceed by splitting $\exppsi$ into a position-dependent part
and a position-independent part
\begin{equation}
\exppsi=\frac{1}{A}\sklamm{1+C(\vek{r})} \label{eq:locvar}
\end{equation}
by imposing the condition that the position average
of $C(\vek{r})$ is zero: $\eklamm{C(\vek{r})}_{\vek{r}}=0$.
Using $\eklamm{1}_{\vek{r}}=1$, we obtain
relation $\eklamm{\exppsi}_{\vek{r}}=A^{-1}=\eklamm{|\Psi(\vek{r})|^2}$.
Introducing the normalized intensity $I=A {\rho}$, 
we can now rewrite the intensity distribution as 

\begin{eqnarray}
P_{\vek{r}}(I) & = & \frac{1}{1+C(\vek{r})} \exp\sklamm{-\frac{I}{1+C(\vek{r})}}
 =
\frac{ e^{-I} }{1-(-C(\vek{r}))} \exp\sklamm{-\frac{I \; 
(-C(\vek{r}))}{(-C(\vek{r}))-1}} 
\nonumber \\ & = &
e^{-I} \; \sum_{n=0}^{\infty} (-1)^n \sklamm{C(\vek{r})}^n L_n(I)
\end{eqnarray}
where in the last step we use the generating function of the Laguerre 
polynomials {$L_n(I)$} \cite{nist}. 
Finally, we perform a position average to obtain, up to second order in $I$,
the normalized intensity distribution
\begin{eqnarray}
P(I){\equiv}\eklamm{P_{\vek{r}}(I)}_{\vek{r}} & = & e^{-I} 
\sklamm{ 1 + \eklamm{C(\vek{r})^2}_{\vek{r}} L_2(I) } \nonumber \\
& {=} & {e^{-I} \sklamm{ 1 + \beta \klamm{1-2 I + \frac{1}{2} I^2}}}
\label{eq:idist}
\end{eqnarray}
with $\beta\equiv\eklamm{C(\vek{r})^2}_{\vek{r}}$. 
In Fig.~\ref{fig:dist_fit} we compare this formula with the numerical{ly}
obtained intensity distributions. 
We see that the numerical data are very well described by a behaviour 
of the form (\ref{eq:idist}), with $\beta$ being the only free parameter.
This supports our claim that 
\textit{for weak interactions, deviations of the intensity
  distribution are universal and depend only on a single parameter}.

\subsection{Semiclassical calculation of $\beta$}
  
The parameter $\beta$ can be numerically obtained by a fitting procedure
and compared with a prediction based on the semiclassical approximation 
to the non-linear Green function $G(\vek{r},\vek{r}',E)$ defined through
\begin{equation}
\label{eq:nlgreen}
\sklamm{E-H-g(\vek{r})\frac{\hbar^2}{m}|G(\vek{r},\vek{r}',E)|^2}
G(\vek{r},\vek{r}',E)=\delta(\vek{r}-\vek{r}') \; .
\end{equation}
In order to understand the connection between the parameter $\beta$
and the nonlinear Green function, we consider first the Green function
$G_{0}$ for the linear system,
\begin{equation}
\label{eq:lgreen}
\sklamm{E-H}G_{0}(\vek{r},\vek{r}',E)=\delta(\vek{r}-\vek{r}'),
\end{equation}
which admits {a} spectral decomposition in terms of the normalized
scattering states $\Psi_{E',\alpha}(\vek{r})$ at energy $E'$ with
incoming channel $\alpha$, given by
\begin{equation}
\label{eq:sgreen}
G^{\pm}_{0}(\vek{r},\vek{r}',E)=\sum_{\alpha}\int dE' 
\frac{\Psi_{E',\alpha}(\vek{r}) \Psi_{E',\alpha}(\vek{r}')^{*}}{E-E'\pm i0^{+}}
\end{equation}
where $0^{+}$ stands for an infinitesimal positive number. 
If we now consider the combination
\begin{equation}
G^{+}_{0}(\vek{r},\vek{r}',E)-G^{-}_{0}(\vek{r},\vek{r}',E)=
-\frac{2}{\pi}\sum_{\alpha}\int dE' \Psi_{E',\alpha}(\vek{r}) 
\Psi_{E',\alpha}(\vek{r}')^{*} \delta(E-E')
\end{equation}
we see that, up to numerical factors, 
\begin{equation}
\langle G^{+}_{0}(\vek{r},\vek{r}',E)-G^{-}_{0}(\vek{r},\vek{r}',E)\rangle_{E} 
\propto \sum_{\alpha}\langle \Psi_{E,\alpha}(\vek{r}) 
\Psi_{E,\alpha}(\vek{r}')^{*}\rangle_{E} {\;.}
\end{equation}
Therefore, the local variance $\exppsi$ can be calculated if we
know the imaginary part of the Green function at $\vek{r}=\vek{r}'$.

Although this construction depends on the fact that $G_{0}$ is the
Green function of a linear operator, our main assumption is that we
can, for weak non-linearities, deform the linear scattering states
into non linear objects such that a spectral decomposition of the
form~\eqref{eq:sgreen} for $G$ holds, at least approximately. Following
the same steps as for the linear case, we conclude that under such
assumptions, the local variance for the interacting case is also
related with the imaginary part of the nonlinear Green function.

Although a closed expression for the non-linear Green function as a
sum over classical paths is not known, it still satisfies a
decomposition of the form
\begin{equation}
\label{eq:slgreen}
G(\vek{r},\vek{r},E)=G^{{\rm zero}}(\vek{r},\vek{r},E)+G^{{\rm long}}(\vek{r},\vek{r},E)
\end{equation}
in terms of zero-length paths joining $\vek{r}$ with $\vek{r}'$ in
zero time, and long paths hitting the boundaries several times. This
decomposition carries over {to} the local variance {which was defined
in Eq.~(\ref{eq:locvar})} as $\exppsi=\frac{1}{A}\sklamm{1+C(\vek{r})}$. 
The contribution from the zero-length paths produces then the uniform 
background $1/A$, while the long paths produce fluctuations around it 
to yield
\begin{equation}
C(r)=\frac{\hbar^2}{m i} \sklamm{G^{{\rm long}}(\vek{r},\vek{r},E)-G^{{\rm long}*}(\vek{r},\vek{r},E)}.
\end{equation}

Finally, the average of $C(\vek{r})^{2}$ is computed within the diagonal
approximation, where different paths are correlated only as long as
they are related by time-reversal symmetry which is assumed to be
weakly broken by the magnetic field. 
In perfect analogy with the derivation of the channel-resolved coherent 
backscattering probability that was calculated in 
Ref.~\cite{weakloc}, we obtain
\begin{eqnarray}
\beta(B,g) & = & {-2\sklamm{\frac{\hbar^2}{m i}}^2 
\eklamm{G^{{\rm long}}(\vek{r},\vek{r}) 
G^{{\rm long}*}(\vek{r},\vek{r})}_{\vek{r}}} \nonumber \\
& = & \frac{\tau_D}{\tau_H} + \frac{\tau_D}{\tau_H} 
\frac{1}{1+\klamm{B/B_{w}}^2} \, \frac{1}{1+\sklamm{2 g j
\frac{\tau_D^2}{\tau_H}\klamm{1+\klamm{B/B_{w}}^2}^{-1}}^2}.
\label{eq:semicl}
\end{eqnarray}
All parameters in this formula are known. 
${\tau}_H={m \Omega / \hbar}$ is the Heisenberg time, and
the dwell time $\tau_D$ as well as the characteristic scale $B_{w}$ 
for the magnetic field are  determined by the classical
dynamics of the system, as shown in Appendix~\ref{sec:classic}.

\begin{figure}[t]
\includegraphics[width=\linewidth]{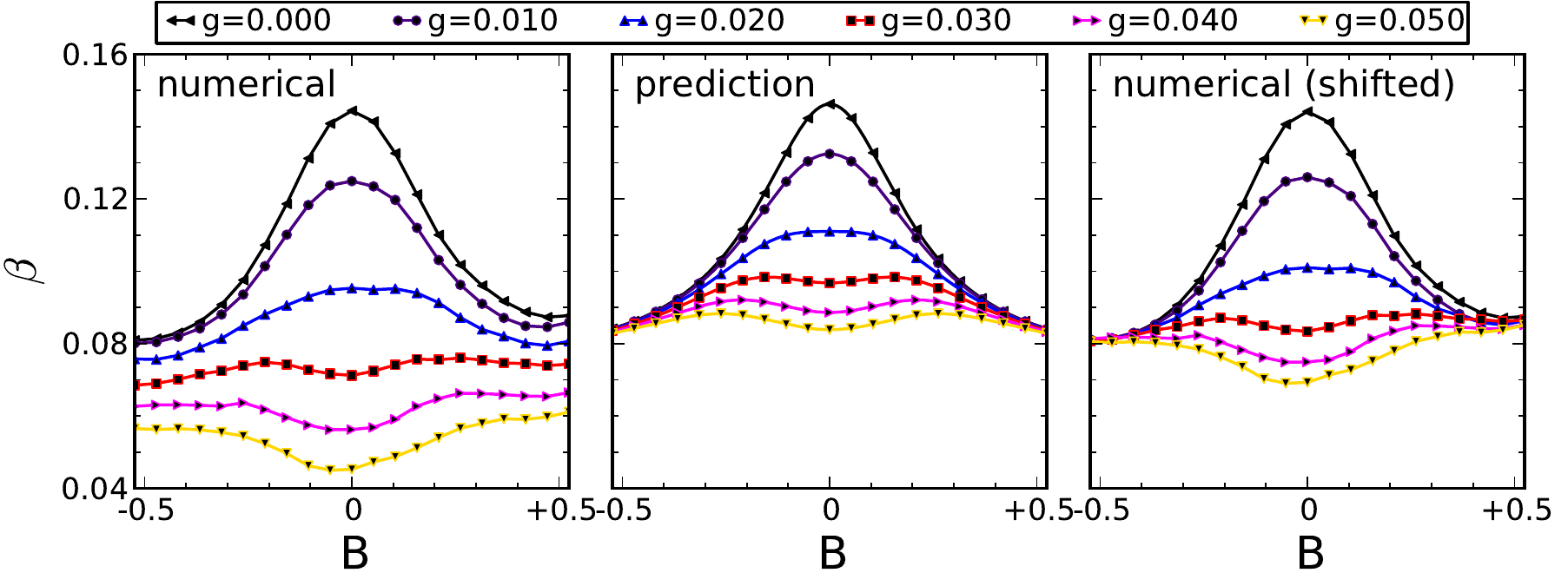}
\caption{\label{fig:fit_all} 
Comparison of the numerically determined values of
$\beta$ (see Fig.~\ref{fig:dist_fit}) in the left panel with the 
prediction from Eq.~\ref{eq:semicl} in the central panel.
Clear deviations are visible, but they have the form of a displacement
$\beta^{0}\equiv\beta^{0}(g)$ that is roughly independent of the magnetic 
field and increases monotonically with $g$.
In the right panel, this displacement $\beta^{0}(g)$ is subtracted from 
the numerical data for $\beta$ such that they match the prediction for 
$B = \pm 0.5$.  }
\end{figure}

Figure~\ref{fig:fit_all} compares the semiclassical 
prediction~\eqref{eq:semicl} with the numerically determined value of 
$\beta$. In the linear case $g=0$ the agreement is very good. 
{In a similar manner as for the channel-resolved retro-reflection
amplitude \cite{weakloc,adist3}, the parameter $\beta$ is enhanced for $B=0$
due to the constructive interference between trajectories that are 
backscattered from $\vek{r}$ to $\vek{r}$ and their time-reversed counterparts.
Finite values of $B$ introduce a dephasing between such trajectories, which
leads to a suppression of the enhancement of the form
$\sim (1 + B^2/ B_w^2)^{-1}$.
Eq.~(\ref{eq:semicl}) predicts that the presence of a repulsive interaction
gives rise to another dephasing mechanism for finite values of $g$,
which, however, is slightly stronger for $B=0$ than for finite $B$ and
can therefore give rise, at finite but small values of $g$, to a local 
minimum of $\beta$ (instead of a maximum) around $B=0$
(see the central panel of Fig.~\ref{fig:fit_all}).
This minimum is found to be slightly more pronounced in the numerically 
determined values for $\beta$ shown in the left panel of 
Fig.~\ref{fig:fit_all}.
As for the case of channel-resolved back-reflection \cite{weakloc},
this discrepancy can be attributed to non-universal short-path contributions,
in particular to self-retraced paths whose contribution to
$\eklamm{G^{{\rm long}}(\vek{r},\vek{r}) 
G^{{\rm long}*}(\vek{r},\vek{r})}_{\vek{r}}$ is doubly counted in
Eq.~(\ref{eq:semicl}).

In addition to this minor discrepancy, we also find more significant
deviations in the form of a {global reduction of the numerical values for 
$\beta$, which is independent of $B$ and increases monotonically with $g$.
Intuitively, this reduction could be explained by the general tendency of
a defocusing nonlinearity to ``smear out'' the intensity distribution within
the billiard, as was already mentioned above in the discussion of
Fig.~\ref{fig:dist}.
Clearly, this tendency would be independent of the presence of a magnetic 
field.
A semiclassical evaluation of this effect, however, is beyond the scope of
this work.
It would, most probably, involve non-linear ladder-type diagrams that modify
expectation values of higher moments of the local intensity as compared to
the linear scattering problem.
As we are, in this work, mainly interested in the dephasing behaviour of
$\beta$ as a function of the magnetic field, we subtract, in the right panel 
of Fig.~\ref{fig:dist}, this global $B$-independent shift from the numerical 
data.
Good agreement is then obtained with the semiclassical prediction.

\section{Conclusion}

\label{sec:conc}

In this contribution we investigated, both numerically and analytically,
the intensity distribution of non-linear scattering states. 
Our approach is based on a mean-field approximation to the fully
interacting problem of an atomic Bose-Einstein condensate, 
where interactions are incorporated by means of a non-linear term 
in the wave equation.
Formally, we therefore expect that similar results hold for other 
kinds of non-linear wave equations, arising, e.g., in nonlinear
optics.

Our main finding is that not only the general features of the
intensity distribution are universally reproduced by a standard Random
Wave Model ansatz, but also that the small deviations from the 
body of the distribution can be understood in this framework by considering 
local Gaussian statistics, in close analogy with the case of linear waves
in classically chaotic geometries. We have finally shown that both the
functional form of the deviations and their theoretical description
by means of local modulations of the mean intensity are governed by 
a single numerical parameter. This parameter has an universal 
contribution originating from long ergodic paths which we were
able to obtain in closed form by means of a semiclassical approach
based on interfering classical trajectories. However, there is also a
contribution that increases monotonically with the nonlinearity and 
is independent of the magnetic field, for which no theoretical
approach is currently available. Once this latter contribution is
identified and  subtracted from the numerical data, we found 
very good agreement of the semiclassical approach with the exact 
numerical calculations.

\begin{acknowledgments}

We would like to thank Josef Michl for valuable assistance in the
evaluation of the semiclassical diagrams.
This work was supported by the DFG {Research Unit} FOR760
``Scattering systems with Complex Dynamics''.

\end{acknowledgments}

\begin{appendix}

\section{Numerical computation of stationary scattering states}
\label{sec:numerics}

In order to numerically solve the non-linear scattering problem,
 we discretize Eq.~\eqref{eq:nl_scatter} using a second-order 
finite-difference approximation \cite{ferry:transport}.
This results in a two-dimensional irregular lattice whose
lattice spacing is chosen such that we have 
approximatively 30 lattice points per wavelength. 
This ensures that the discretization error is negligible.
The artificial gauge field $\vek{A}(\vek{r})$ is incorporated
through a Peierls phase \cite{peierls@1933}.
 
The interaction strength $g(\vek{r})$ is assumed to be
constant throughout the billiard but adiabatically ramped off
inside the leads as explained in Ref.~\cite{paul:bigrev}.
Therefore, the effects of the leads can be described, 
as in the linear case, by self-energies which provide the correct 
outgoing boundary conditions.
This allows us to restrict the simulation to a finite spatial region.
This procedure is analogous to the approach used in the recursive 
Green function method \cite{datta:meso,lee@fisher@rg}.

\begin{figure}
 \begin{minipage}[c]{0.48\linewidth}
  \includegraphics[width=1\linewidth]{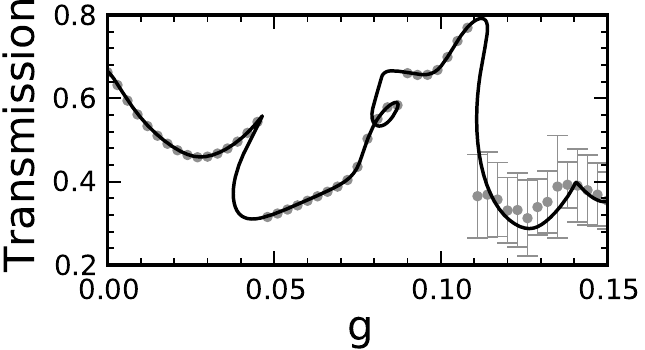}%
 \end{minipage}\hfill
 \begin{minipage}[c]{0.48\linewidth}
\caption{\label{fig:gtrack} The black curve shows the transmission obtained 
using the continuation method.
Results obtained through a time-dependent population of the 
billiard using Eq.~(\ref{eq:gp_tp}) are shown in gray. 
For large $g$, no dynamically stable stationary solution exists.
}
\end{minipage}
\end{figure}

The complex wavefunction $\Psi(\vek{r})$ is now represented
by a $2\mathcal{N}$-dimensional real vector{, with $\mathcal{N}$} the 
number of lattice points{.}
Defining 
\begin{equation}
F: \mathbb{R}^{2\mathcal{N}} \to \mathbb{R}^{2\mathcal{N}}
\quad\quad
\Psi(\vek{r}) \mapsto \sklamm{\mu-H}\Psi(\vek{r})
-g(\vek{r})\frac{\hbar^2}{m}|\Psi(\vek{r})|^2\Psi(\vek{r}) - S(\vek{r}) \,,
\end{equation}
we have to seek for solutions of $F(\Psi)=0$. 
This is done with Newton's iteration \cite{wright:opt}
$\Psi_{k+1}=\Psi_{k}-\klamm{\mathcal{D}F}^{-1} F(\Psi_k)$ 
which converges to a zero of $F$ provided that the starting vector 
$\Psi_0$ is suitabl{y} chosen.
This choice is a non-trivial matter.
Using $g$ as an additional free parameter {--- i.e.,} 
$g(\vek{r}){\equiv}g\, g_0(\vek{r})$ with $g\in\mathbb{R}$ and
$g_0(\vek{r})=1$ for $\vek{r}$ inside the billiard {---} 
we re{-}interpret $F{\equiv}F(\Psi(\vek{r}),g)$ as
a function 
$F: \mathbb{R}^{2\mathcal{N}} \times \mathbb{R} \to \mathbb{R}^{2\mathcal{N}}$.
Neglecting critical points{,} the set $F^{-1}(0)$ is a one-dimensional 
manifold which can be traced  by a continuation method 
\cite{seydelchaos,wright:opt}
yielding the manifold as a parametric function $s \mapsto (\Psi(s),g(s))$
of the arclength $s$. 
An example of such a calculation is shown in Fig.~\ref{fig:gtrack}. 

A prominent feature of non-linear wave equations is their potential 
multi-stability, i.e., the fact
that they can support multiple solutions for a fixed value of $g$. 
In such a situation, the state that would be populated in an 
experiment depends on the history of the system.
Here, we always use the the first solution found by the continuation 
method.
This choice mimics the time-dependent population of the billiard that
would be obtained from integrating Eq.~\eqref{eq:gp_tp} in the presence
of an adiabatically slow increase of the source amplitude.

Additional details of the numerical methods can be found in 
{Refs.~\cite{weakloc,Hartmann}}.

\section{Analysis of the classical dynamics}
\label{sec:classic}

The parameters $\tau_D$ and $B_{w}$ in Eq.~\eqref{eq:semicl} 
can be determined using classical simulations. 
To this end, classical trajectories inside the billiard
are calculated using a ray-tracing algorithm.
The trajectories start in the left lead at a given longitudinal position
$x$ with a given total momentum $p = \sqrt{p_x^2 + p_y^2}$, while
the transverse coordinate $y$ and the associated component $p_y$ of the
momentum are randomly selected in a uniform manner.
The simulation is continued until the trajectory leaves 
through one of the leads.

\begin{figure}
\includegraphics[width=0.49\linewidth]{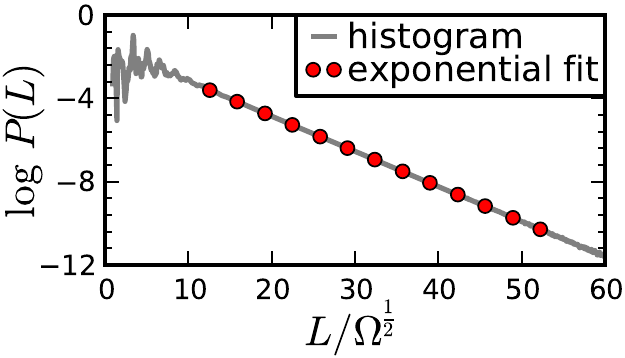}
\hfill
\includegraphics[width=0.49\linewidth]{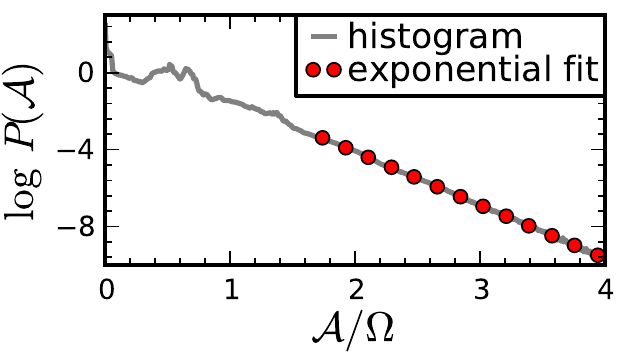}
\caption{\label{fig:cladyn} 
  Probability distributions
  of path lengths $L$ (left) and directed areas $\mathcal{A}$ (right) of classical trajectories
  inside the billiard that is shown in Fig.~\ref{fig:billiard}. 
  An exponential function
  is fitted (dots) onto both distributions after a short transient region.
}
\end{figure}

The time $t_{\gamma}$ spent inside the cavity follows an exponential 
distribution
\begin{equation}
P(t_{\gamma})=\tau_D e^{-t_{\gamma}/\tau_D}
\end{equation}
where $\tau_{D}$ is the classical dwell time. 
Thus, an exponential fit (shown in Fig.~\ref{fig:cladyn}) of the 
numerically obtained path-length distribution yields the 
classical dwell time $\tau_D$.
Its numerical value is, in our units, given by the average 
population $j_{in} \tau_D \simeq 241$.

A central limit ansatz results in {the} Gaussian distribution 
\begin{equation}
P(t_{\gamma},{\mathcal{A}})=\frac{1}{\sqrt{2\pi t_{\gamma} {\eta}}} 
{\; \exp}\left(-\frac{{\mathcal{A}}^2}{2 t_{\gamma} {\eta}}\right)
\end{equation}
for the directed areas $\mathcal{A}$ for paths of a given time 
$t_{\gamma}$.
Here, ${\eta}$ is a geometry-dependent parameter that 
can be determined by evaluating the total distribution of 
$\mathcal{A}$,
\begin{equation}
P({\mathcal{A}})=\tau_D \int_0^{+\infty} P(t_{\gamma},{\mathcal{A}}) e^{-t_{\gamma}/\tau_D} dt
=\frac{1}{\sqrt{2 {\eta} t_D}} \exp\sklamm{-\sqrt{\frac{2}{{\eta} \tau_D}} \norm{{\mathcal{A}}}}
\; .
\end{equation}
This is also an exponential distribution, 
and thus an exponential fit can be used to compute ${\eta}$ 
as shown in Fig.~\ref{fig:cladyn}.
The parameter $B_{w}$ is now finally determined as
\begin{equation}
  B_{w} = \frac{\hbar}{q} \frac{1}{\sqrt{2 {\eta} \tau_D }} {\,.}
\end{equation}
{We numerically find} $B_{w}=0.22 \, B_0$ in our units.
Additional details are given in Ref.~\cite{weakloc}.

\end{appendix}

\end{document}